\documentclass[twocolumn,showpacs,showkeys,preprintnumbers,amssymb,amsmath,aps,superscriptaddress,prb]{revtex4-1}
\usepackage{graphicx}% Include figure files
\usepackage{dcolumn}% Align table columns on decimal point
\usepackage{bm}% bold math
\usepackage{epsfig}

\begin{document}

%\preprint{APS/123-QED}

\title{Non-perturbative linked-cluster expansions for the trimerized ground state of the spin-one Kagome Heisenberg model}

\author{Dominik Ixert}
\email{dominik.ixert@tu-dortmund.de}
\affiliation{Lehrstuhl f\"ur Theoretische Physik I, Otto-Hahn-Strasse 4, TU Dortmund, 44221 Dortmund, Germany}

\author{Tobias Tischler}
\email{tobias.tischler@tu-dortmund.de}
\affiliation{Lehrstuhl f\"ur Theoretische Physik I, Otto-Hahn-Strasse 4, TU Dortmund, 44221 Dortmund, Germany}

\author{Kai P. Schmidt}
\email{kai.schmidt@tu-dortmund.de}
\affiliation{Lehrstuhl f\"ur Theoretische Physik I, Otto-Hahn-Strasse 4, TU Dortmund, 44221 Dortmund, Germany}

\date{\today}

\begin{abstract}
We use non-perturbative linked-cluster expansions to determine the ground-state energy per site of the spin-one Heisenberg model on the kagome lattice. To this end, a parameter is introduced allowing to interpolate between a fully trimerized state and the isotropic model. The ground-state energy per site of the full graph decomposition up to graphs of six triangles (18 spins) displays a complex behaviour as a function of this parameter close to the isotropic model which we attribute to divergencies of partial series in the graph expansion of quasi-1d unfrustrated chain graphs. More concretely, these divergencies can be traced back to a quantum critical point of the one-dimensional unfrustrated chain of coupled triangles. Interestingly, the reorganization of the non-perturbative linked-cluster expansion in terms of clusters with enhanced symmetry yields a ground-state energy per site of the isotropic two-dimensional model being in quantitative agreement with other numerical approaches in favor of a spontaneous trimerization of the system. Our findings are of general importance for any non-perturbative linked-cluster expansion on geometrically frustrated systems.
\end{abstract}


\maketitle

\section{Introduction} 
The search for exotic phases of correlated quantum matter is one of the most attractive issues
 in physics, since one expects fascinating types of collective quantum behaviour hosting unconventional physical properties of potential interest for future technologies. To this end, the interplay of frustrated interactions and strong quantum fluctuations is believed to represent the key knobes one can tune to stabilize unconventional types of quantum order.  

In this context the physics of the quantum spin-1/2 Heisenberg model on the highly frustrated two-dimensional kagome lattice plays a paradigmatic role and a great amount of research has been performed over the last decades. The long history of that problem directly reflects the enormous complexity of such highly frustrated quantum magnets. Due to recent numerical investigations \cite{Yan11,Depenbrock12}, the nature of the ground state is most likely a quantum spin liquid exhibiting exotic topological order. Elementary excitations above the topologically-ordered ground state are therefore neither bosons are fermions, but anyons carrying fractional spin quantum numbers. 

The physics of the spin-one cousin of the Heisenberg model on the kagome lattice has been focused on recently \cite{Hida00,Yao10,Nishimoto14,Li14,Changlani15,Liu15,Li15}. Interestingly, several works point towards a spontanenous trimerization of the system \cite{Changlani15,Liu15,Li15}, i.e.~the symmetry between up- and down-triangles is broken spontaneously and a singlet ground state with no long-range magnetic order is found. As a consequence of breaking this discrete symmetry, the spin-one Heisenberg model is expected to be gapped. This is indeed in agreement with recent density matrix renormalization group (DMRG) calculations pointing towards a finite spin gap \cite{Changlani15}. Nevertheless, there are also competing numerical works which argue in favor of different ordered or disordered ground states \cite{Hida00,Yao10,Li14}, and it is therefore important to further clarify the physical properties of the spin-one Heisenberg model on the kagome lattice. On the experimental side, there are also several materials like m-MPYNN$\cdot$BF$_4$ \cite{Awaga94,Wada97,Watanabe98,Matsushita10} or Ni$_3$V$_2$O$_8$ \cite{Lawes04} which seem to correspond microscopically to spin-one systems on the kagome lattice. 

If the spin-one Heisenberg model on the kagome lattice displays spontaneous trimerization, then it should be possible to deform the ground state of the Heisenberg model adiabatically up to the fully trimerized limit of isolated triangles without encountering any quantum phase transition. In this work we check this scenario by calculating the ground-state energy per site in the thermodynamic limit using non-perturbative linked-cluster expansions (NLCEs) \cite{Rigol06,Rigol07_1,Rigol07_2,yang11,Tang13,coester15}. We set up a full graph decomposition in terms of up-triangles of the kagome lattice and we perform exact diagonalization (ED) on these graphs. We find a surprisingly complex convergence behaviour of the ground-state energy per site. This is shown to originate from a quantum critical point present in the one-dimensional subsystem of coupled triangles. The traces of this quantum critical point can be consistently avoided by reorganizing the NLCE of the two-dimensional problem in terms of more symmetric clusters. This reorganized NLCE displays a strongly improved convergence and our results are fully consistent with a trimerized ground state of the spin-one Heisenberg model on the kagome lattice. On the technical side, the appearance of divergent partial series in non-perturbative linked-cluster expansions is expected to be a generic feature in geometrically frustrated systems.

The outline of this paper is as follows. In Sect.~\ref{Sect:model} we introduce the trimerized version of the spin-one Heisenberg model on the kagome lattice. Next, we give technical details on the non-perturbative linked-cluster expansions performed in this work in Sect.~\ref{Sect:nlce} and we present all physical results in Sect.~\ref{Sect:results}. Finally, the most important outcomes of our work are briefly concluded in Sect.~\ref{Sect:conclusions}.

\section{Model}
\label{Sect:model}
%
%Figure 1: Illustration Lattice / Model /Effective lattice
%%%%%%%%%%%%%%%%%%%%%%%%%%%%%%%%%%%%%%%%%%%%%%%%%%%%%%%%%%%%%%%%%%%%%%%%%%%%%%%%%%%%%%%%%%%%%
\begin{figure}
\begin{center}
\epsfig{file=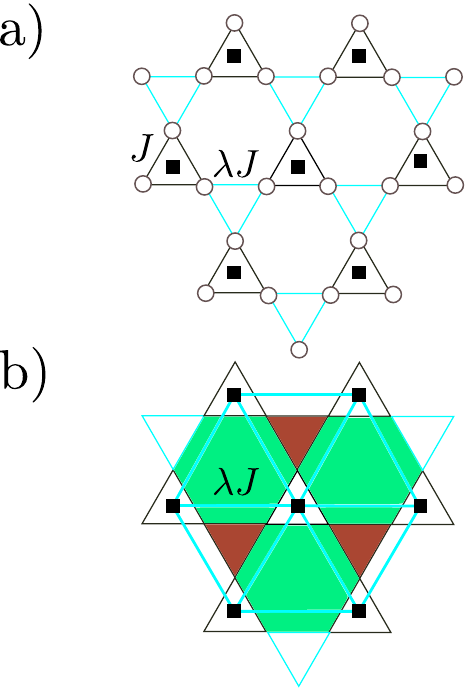, width=0.65\columnwidth}
\end{center}
\caption{(Color online) a) A schematic illustration of the spin-one Heisenberg model on the kagome lattice. Spins are located on the vertices of the lattice highlighted as circles. The Heisenberg exchange on up- (down-) triangles is $J$ ($\lambda J$) and illustrated as black (gray/cyan) lines. The black boxes mark the center of up-triangles. b) The effective triangular lattice of up-triangles is shown. Note that up- and down-pointing triangles (light/green and dark/red areas) in the effective lattice are topologically distinct, since they correspond to different plaquettes (hexagons and down-triangles) in the original kagome lattice.}
\label{fig:kagome}
\end{figure}
%%%%%%%%%%%%%%%%%%%%%%%%%%%%%%%%%%%%%%%%%%%%%%%%%%%%%%%%%%%%%%%%%%%%%%%%%%%%%%%%%%%%%%%%%%%%%%%

Let us consider a trimerized version of the spin-one Heisenberg model on the kagome lattice as illustrated in Fig.~\ref{fig:kagome}. The microscopic Hamiltonian is given by
\begin{equation}
\mathcal{H} = J \sum_{\langle i,j\rangle\in\Delta } {\mathbf S}_{i}\cdot{\mathbf S}_{j} +  \lambda J \sum_{\langle i,j\rangle\in\nabla } {\mathbf S}_{i}\cdot{\mathbf S}_{j}\quad , 
\label{eq:ham}
\end{equation}
so that the first (second) sum runs over all nearest neighbor sites on up-triangles (down-triangles). The real parameter $\lambda\in\left[0,1\right]$ allows an interpolation between the limit of isolated up-triangles $\lambda=0$ and the isotropic Heisenberg model at $\lambda=1$. In the following we focus on an antiferromagnetic exchange constant $J>0$ and we set $J=1$.

For $\lambda=0$, the fully trimerized system has a unique singlet ground state which is given as the product state of singlets on individual up-triangles. Indeed, in contrast to its spin-1/2 cousin where an isolated antiferromagnetic triangle is four-fold degenerate, the addition of three spin-ones allows to form a unique singlet which has the lowest energy. All excited states at $\lambda=0$ are gapped and can be understood via the remaining 26 states of an isolated triangle and their magnetic quantum numbers. 

If the isotropic model at $\lambda=1$ displays a spontaneous trimerization, then one expects that ground states at $\lambda=0$ and $\lambda=1$ are adiabatically connected and, consequently, no energy gap should close in the whole interval $\lambda\in\left[0,1\right]$. This scenario we explore in the following by applying NLCEs which we explain next.

\section{NLCE}
\label{Sect:nlce}

The essential idea of NLCEs is to exploit the linked-cluster theorem so that actual 
numerical calculations are done on finite linked clusters, but the final results are valid 
directly in the thermodynamic limit. Generically, NLCEs consists of three steps: i) choosing and generating the families of clusters or topologically distinct graphs used in the linked-cluster expansion, ii) performing numerical calculations on graphs extracting the physical quantities of interest, and iii) determing the reduced contributions specific to each graph and embed these contributions into the infinite lattice.

The details of the choice of graphs (i) will be given below. Concerning ii), we are using ED with the Lanczos algorithm to determine the ground-state energy $E^{\mathcal{G}_\nu}_0$ on each graph and for each value of $\lambda$. Apart from the exponential increase of graphs with the number of sites $N$, the memory needed for ED is the limiting factor of NLCEs. Practically, we consider graphs with at most 18 spin-ones in our calculations. 

In the third step iii), the ground-state energy per site $\epsilon_0$ in the thermodynamic limit is expressed as
%%%%%%%%%%%%%%%%%%%%%%%%%%%%%%%%%%%%%%%%%%%%%%%%%%%%%%%%%%%%%%%%%%%%%%%%%%%%%%%%%%%%%%%%%%%%%
\begin{equation}
\epsilon_0  = \sum_{j} \nu^{\mathcal{G}_j} \, \epsilon_{0,{\rm red}}^{\,\mathcal{G}_j} \quad , 
\label{eq:e0}
\end{equation}
%%%%%%%%%%%%%%%%%%%%%%%%%%%%%%%%%%%%%%%%%%%%%%%%%%%%%%%%%%%%%%%%%%%%%%%%%%%%%%%%%%%%%%%%%%%%%
where the sum runs over all linked graphs $\mathcal{G}_j$. The integer number $\nu^{\mathcal{G}_j}$ is the so-called embedding factor specifying the combinatorical number how often graph $\mathcal{G}_j$ can be embedded into the infinite lattice. The reduced contribution $\epsilon_{0,{\rm red}}^{\,\mathcal{G}_j}$ specific to graph $\mathcal{G}_j$ results from subtracting all contributions of subgraphs in order to avoid double counting. One has
%%%%%%%%%%%%%%%%%%%%%%%%%%%%%%%%%%%%%%%%%%%%%%%%%%%%%%%%%%%%%%%%%%%%%%%%%%%%%%%%%%%%%%%%%%%%%
\begin{equation}
 \epsilon_{0,{\rm red}}^{\,\mathcal{G}_j} = E_0 ^{\mathcal{G}_j} - \sum_{\mathcal{G}_{j'}\subset \mathcal{G}_{j}} \nu^{\mathcal{G}_{j'}}\, \epsilon_{0,{\rm red}}^{\,\mathcal{G}_{j}} \quad . 
\label{eq:e0_tilde}
\end{equation}
%%%%%%%%%%%%%%%%%%%%%%%%%%%%%%%%%%%%%%%%%%%%%%%%%%%%%%%%%%%%%%%%%%%%%%%%%%%%%%%%%%%%%%%%%%%%%
In the following we detail the two different types of \mbox{NLCEs} we performed for the spin-one Heisenberg model on the kagome lattice.
%
% Subsection full-graph expansion
%%%%%%%%%%%%%%%%%%%%%%%%%%%%%%%%%%%%%%%%%%%%%%%%%%%%%%%%%%%%%%%%%%%%%%%%%%%%%%%%%%%%%%%%%%%%%
\subsection{Full-graph decomposition in triangles}
\label{SSect:graphs_full}
We are aiming at describing the trimerized ground state for $\lambda\in [0,1]$. Is is therefore reasonable to set up a full-graph expansion {\it not} in terms of individual spin sites, but in terms of up-triangles (or equivalently of down-triangles) as effective supersites. The effective lattice of these supersites corresponds to a triangular lattice as illustrated in the lower panel of Fig.~\ref{fig:kagome}. The up- and down-triangles of the effective triangles are topologically distinct, since they correspond to hexagons and triangles in the original kagome lattice. Respecting this property, we performed a full graph expansion in the effective lattice for all graphs up to $N_{\rm tr}=6$ supersites. i.e.~6 up-triangles in the kagome lattice consisting of 18 spins one. 
%
%Figure 2: Illustration graphs full
%%%%%%%%%%%%%%%%%%%%%%%%%%%%%%%%%%%%%%%%%%%%%%%%%%%%%%%%%%%%%%%%%%%%%%%%%%%%%%%%%%%%%%%%%%%%%
\begin{figure}
\begin{center}
\epsfig{file=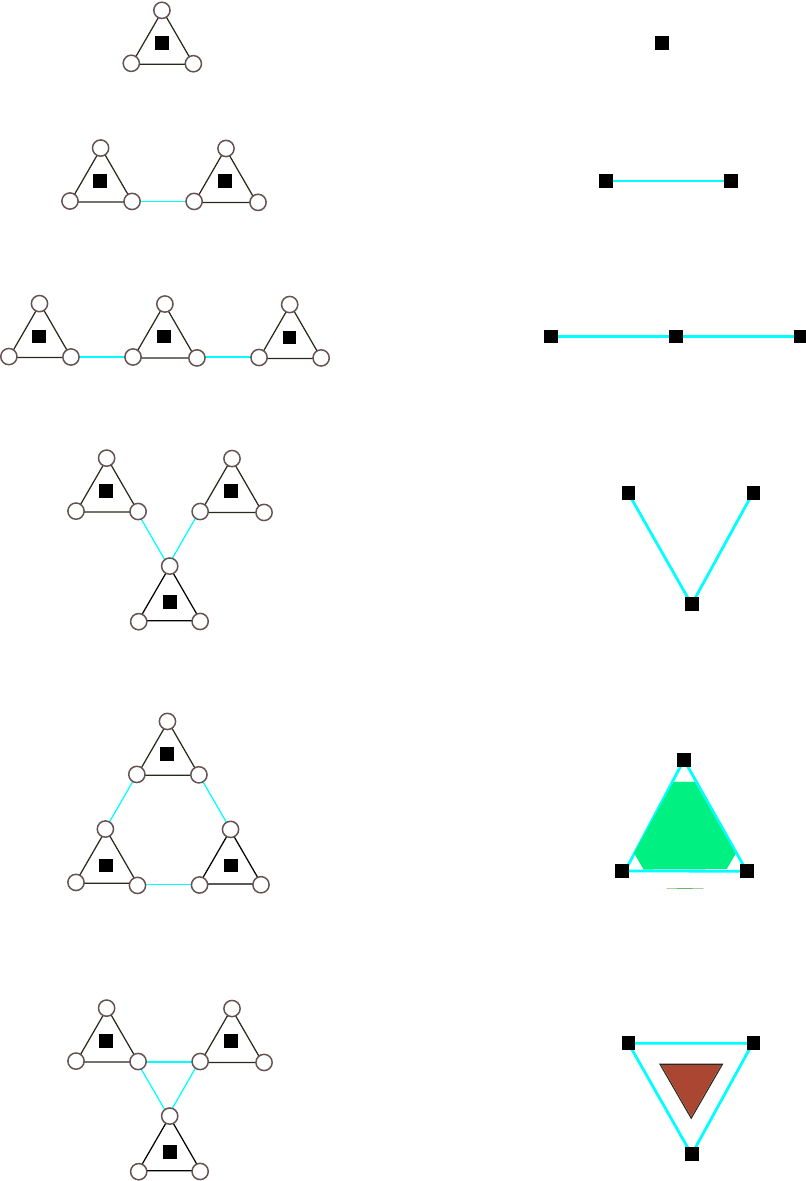, width=\columnwidth}
\end{center}
\caption{(Color online) Illustration of all graphs used in the full graph expansion up to $N_{\rm tr}=3$ triangles. The left panel corresponds to the original kagome lattice while the right panel shows the same graphs in the effective lattice where up-triangles (shown in black) are replaced by supersites (black boxes).}
\label{fig:NLCE_full}
\end{figure}
%%%%%%%%%%%%%%%%%%%%%%%%%%%%%%%%%%%%%%%%%%%%%%%%%%%%%%%%%%%%%%%%%%%%%%%%%%%%%%%%%%%%%%%%%%%%%%%

All graphs up to $N_{\rm tr}=3$ are illustrated in Fig.~\ref{fig:NLCE_full} in the original kagome lattice (left panel) as well as in the effective triangular lattice (right panel). In total, there are 1, 1, 4, 9, 46, 228 topologically distinct graphs for $N_{\rm tr}$ from 1 to 6.  Finally, we have determined the ground-state energy per site $\epsilon_0^{(N_{\rm tr})}$ taking into account all graphs up to $N_{\rm tr}\in\{1,\ldots,6\}$. 
%
% Subsection reorganized expansion
%%%%%%%%%%%%%%%%%%%%%%%%%%%%%%%%%%%%%%%%%%%%%%%%%%%%%%%%%%%%%%%%%%%%%%%%%%%%%%%%%%%%%%%%%%%%%
\subsection{Reorganized graph expansion}
\label{SSect:graphs_opt}
%%%%%%%%%%%%%%%%%%%%%%%%%%%%%%%%%%%%%%%%%%%%%%%%%%%%%%%%%%%%%%%%%%%%%%%%%%%%%%%%%%%%%%%%%%%%%
Next we introduce a reorganized graph expansion which turns out to be well suited for the spin-one Kagome Heisenberg model as detailed below. The essential idea is to restrict the calculation to only a small subset of clusters which, however, are highly symmetric.

\begin{figure}
\begin{center}
\epsfig{file=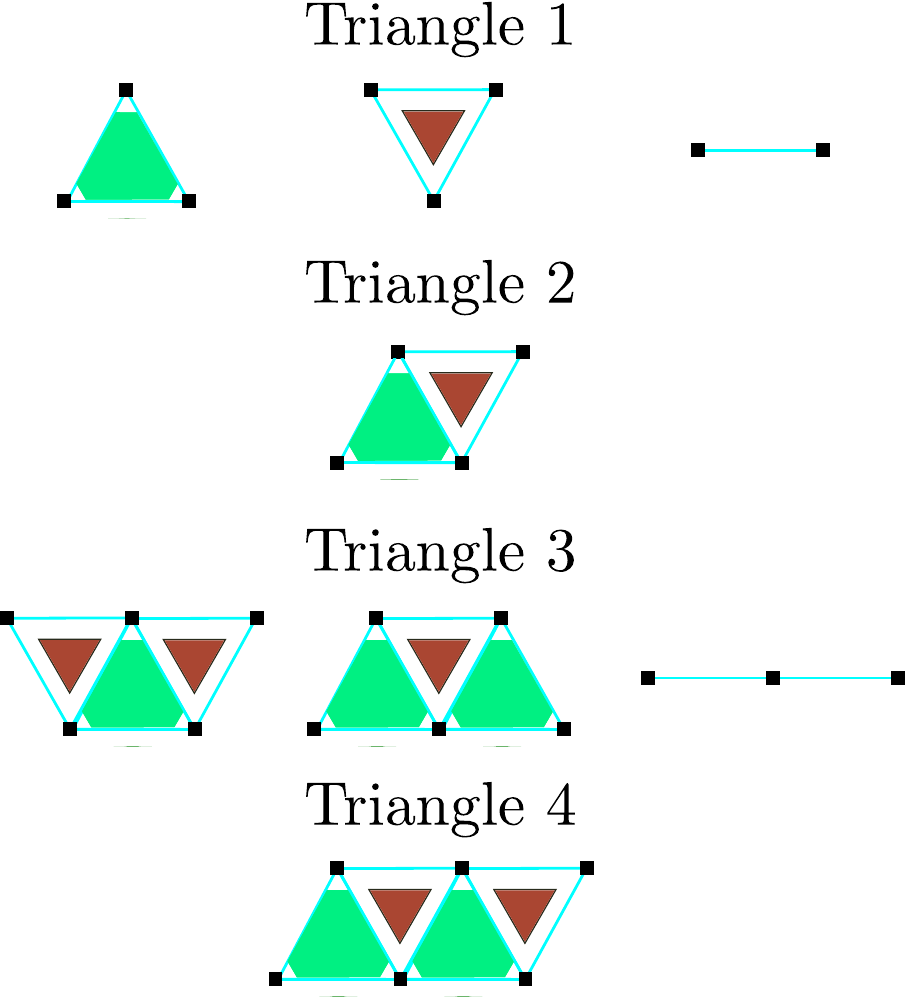, width=0.9\columnwidth}
\end{center}
\caption{(Color online) Illustration of all graphs used in the reorganized expansion except the one consisting just of a single supersite. The idea is to use the two inequivalent triangles of the effective triangular lattice as the building blocks of a NLCE calculation.}
\label{fig:NLCE_triangle}
\end{figure}

In terms of the effective triangular lattice as shown in Fig.~\ref{fig:kagome}b, the important building blocks of this lattice are the two inequivalent triangles as displayed in the upper panel of Fig.~\ref{fig:NLCE_triangle}. We use these two objects as the central blocks for the reorganized graph expansion. In the simplest approximation $\epsilon_0^{\rm tr(1)}$ we restrict the NLCE calculation to these two clusters plus the two-site chain graph which needs to be taken into account in order to avoid double counting. Note that one has to consider the single supersite graph which therefore corresponds formally to $\epsilon_0^{\rm tr(0)}$. This scheme is then extended by adding only clusters made (symmetrically) of effective up- and down-triangles so that neighboring triangles always share edges and not corners as shown in Fig.~\ref{fig:NLCE_triangle}. In the following we numerate this reorganized graph expansion by the total number of triangles involved so that we determined $\epsilon_0^{{\rm tr}(n)}$ with $n\in\{0,1,2,3,4\}$. Finally, one can further reorganize the graph expansion to even more symmetric clusters made of effective plaquettes where a plaquette consists of one effective down- and up-triangle. This expansion is performed exactly when restricting to even $n$.  
%
%Figure 3: Illustration graphs reorganized
%%%%%%%%%%%%%%%%%%%%%%%%%%%%%%%%%%%%%%%%%%%%%%%%%%%%%%%%%%%%%%%%%%%%%%%%%%%%%%%%%%%%%%%%%%%%%
%%%%%%%%%%%%%%%%%%%%%%%%%%%%%%%%%%%%%%%%%%%%%%%%%%%%%%%%%%%%%%%%%%%%%%%%%%%%%%%%%%%%%%%%%%%%%%%

\subsection{Extrapolation}
\label{SSect:extrapolation}
Within the NLCE we get a finite series for the ground-state energy per site $\epsilon_0$ in the thermodynamic limit. For the limit $N_{\rm tr} \to \infty$ one expects to obtain the exact result. In this context the $\epsilon_0^{(N_{\rm tr})}$ corresponds to the partial sum of all orders up to $N_{\rm tr}$. To increase the convergence of the series there is a powerful technique called ``series acceleration''. 
For our problem the Wynn algorithm~\cite{Guttmann89,Tang13} is of particular interest. It takes the form
\begin{align}
	e_n^{(-1)} &= 0 \\
	e_n^{(0)} &= \epsilon_0^{(n)}\\
	e_n^{(k)} &= e_{n+1}^{(k-2)} + \frac{1}{e_{n+1}^{(k-1)} - e_{n}^{(k-1)}} \; .
\end{align}
The parameter $n$ runs from $1$ to $N_{\rm tr}$ in the initial step $e_n^{(0)}$ and is lowered by $1$ in each Wynn step. In practize, one can also restrict to certain orders, e.g.~one could only take into account even $N_{\rm tr}$ or one omits lowest-order contributions by setting $e_n^{(0)} = \epsilon_0^{(n+1)}$. For odd $k$ the entries usually diverge, whereas they converge for even $k$. At the end the sequence $e_{N-2l}^{2l}$ (for $l=0,1,2,\dots$) converges to $\epsilon_{0}$.  

%
%Figure 4: Illustration one-dimensional chain
%%%%%%%%%%%%%%%%%%%%%%%%%%%%%%%%%%%%%%%%%%%%%%%%%%%%%%%%%%%%%%%%%%%%%%%%%%%%%%%%%%%%%%%%%%%%%
\begin{figure}
\begin{center}
\epsfig{file=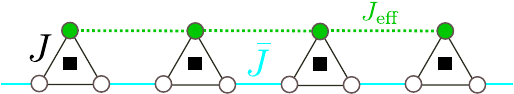, width=0.95\columnwidth}
\end{center}
\caption{(Color online) Illustration of the one-dimensional spin-one triangle chain. Spins are located on the vertices of the lattice highlighted as black circles. The Heisenberg exchange on up-triangles is $J$ (black lines) while the unfrustrated Heisenberg exchange is $\bar{J}$ between triangles forming dimers (shown as gray/cyan lines). In the limit $J\rightarrow 0$ one can derive an effective spin-one Heisenberg chain in second-order perturbation theory in $J/\bar{J}$ for the spin ones building the tips of the black triangles (dark/green circles). The effective Heisenberg exchange $J_{\rm eff}$ is antiferromagnetic and illustrated as dotted lines.}
\label{fig:triangle_chain}
\end{figure}
%%%%%%%%%%%%%%%%%%%%%%%%%%%%%%%%%%%%%%%%%%%%%%%%%%%%%%%%%%%%%%%%%%%%%%%%%%%%%%%%%%%%%%%%%%%%%%%

\section{Results}
\label{Sect:results}

We start the discussion with a one-dimensional triangle chain which represents a one-dimensional subsystem of the trimerized Kagome lattice. This model is essentially unfrustrated (in terms of the effective triangles) and, in addition, the full graph expansion becomes very simple due to the one-dimensional nature. This system is shown to possess a quantum critical point which has severe consequences for the full-graph expansion of the two-dimensional trimerized Kagome lattice presented afterwards. Finally, we discuss the physical properties of the reorganized NLCE which displays a very good convergence behaviour.

%Figure: GSE Triangular chain
%%%%%%%%%%%%%%%%%%%%%%%%%%%%%%%%%%%%%%%%%%%%%%%%%%%%%%%%%%%%%%%%%%%%%%%%%%%%%%%%%%%%%%%%%%%%%
\begin{figure}
\begin{center}
\epsfig{file=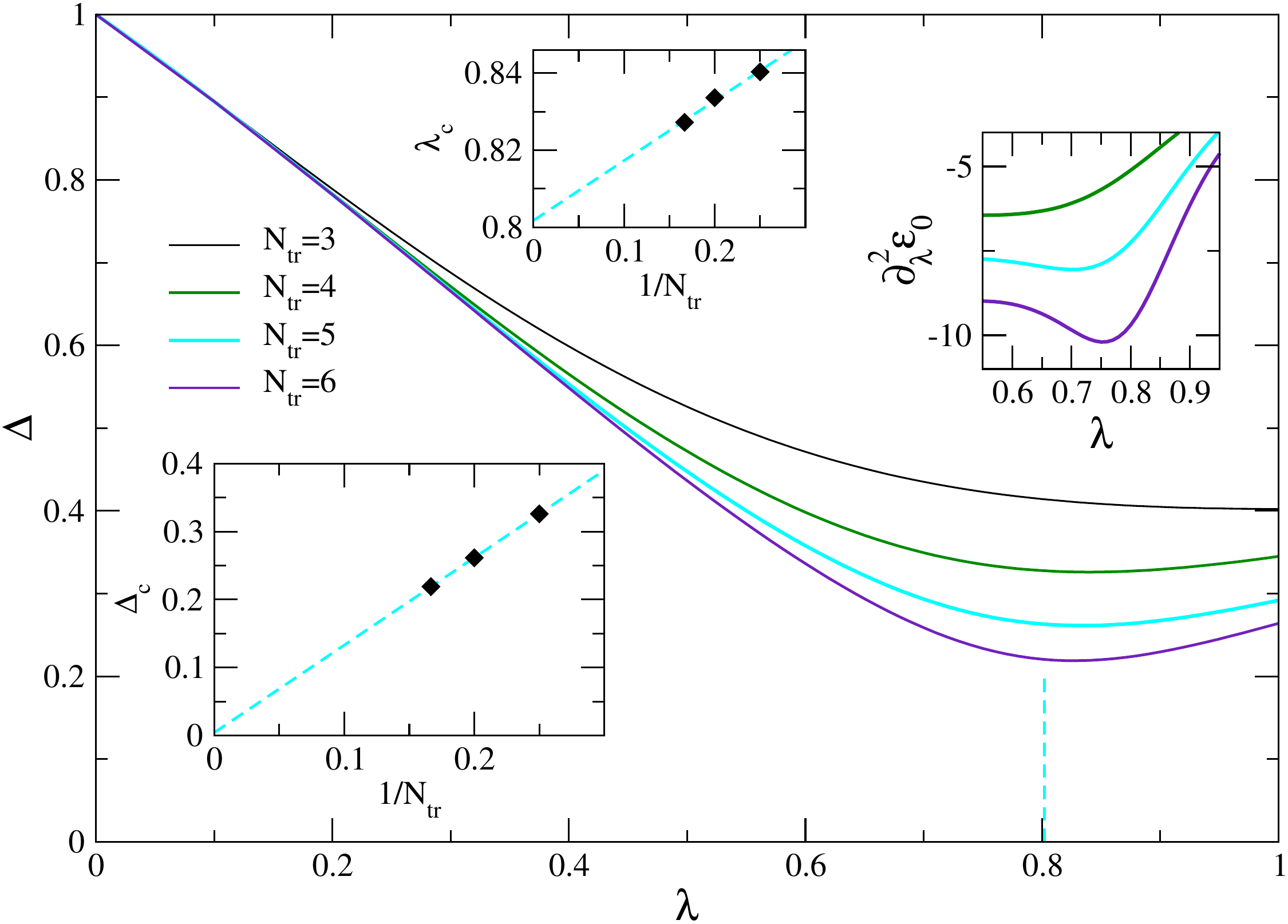, width=\columnwidth}
\end{center}
\caption{(Color online) Spin gap $\Delta$ as a function of $\lambda$ for the one-dimensional triangle chain with periodic boundary conditions, $J=1$, and number of triangles $N_{\rm tr}$. The dashed vertical line marks the location of the quantum critical point in the thermodynamic limit. {\it Inset left}: Minimal spin gap $\Delta_{\rm c}$ as a function of $1/N_{\rm tr}$. The dashed line represents a linear fit through the data points signaling a quantum phase transition in the thermodynamic limit. {\it Inset middle}: Ratio $\lambda_{\rm c}$ corresponding to the minimal value of the spin gap $\Delta_{\rm c}$ as a function of $1/N_{\rm tr}$. The dashed line represents a linear fit through the data points signaling the location of the quantum critical point in the thermodynamic limit. {\it Inset right}: Second derivative $\partial^2_\lambda \epsilon_0$ as a function of $\lambda$.}
\label{fig:triangle_chain_pbc}
\end{figure}
%%%%%%%%%%%%%%%%%%%%%%%%%%%%%%%%%%%%%%%%%%%%%%%%%%%%%%%%%%%%%%%%%%%%%%%%%%%%%%%%%%%%%%%%%%%%%%%

\subsection{One-dimensional triangle chain}
\label{SSect:results_1d}
We consider the one-dimensional spin-one triangle chain illustrated in Fig.~\ref{fig:triangle_chain}. The Hamiltonian is given by
\begin{equation}
\mathcal{H} = J \sum_{\langle i,j\rangle\in\Delta } {\mathbf S}_{i}\cdot{\mathbf S}_{j} +  \bar{J} \sum_{\langle i,j\rangle\in{\rm dimer}} {\mathbf S}_{i}\cdot{\mathbf S}_{j}\quad , 
\label{eq:ham_chain}
\end{equation}
so that the first sum runs over all nearest neighbor sites on triangles while the second sum contains all nearest neighbor sites between adjacent triangles forming spin-one dimers. In the limit $\bar{J}\equiv\lambda J \rightarrow 0$, one is close to the trimerized limit as for the two-dimensional kagome counterpart, while in the limit $J\rightarrow 0$ with $\bar{J}$ finite the system realizes singlets on dimers between triangles while there are an extensive number of isolated spin ones building the tips of the up-triangles. If one performs second-order degenerate perturbation theory in this extensive subspace of isolated spins, one obtains up to an irrelevant constant $E_0^{\rm eff}$ the Hamiltonian of an effective antiferromagnetic spin-one Heisenberg chain 
\begin{equation}
 \mathcal{H}_{\rm eff}=E_0^{\rm eff}+ J_{\rm eff} \sum_{\langle i,j\rangle} {\mathbf S}_i\cdot {\mathbf S}_j \quad ,
\end{equation}
where \mbox{$J_{\rm eff}=\frac{4}{3} J^2/\bar{J}$}. This systems is known to realize a symmetry-protected topologically-ordered ground state with a finite Haldane gap \cite{Haldane83,Haldane83b,Berg08,Gu09,Pollmann12}. One therefore expects at least one phase transition when varying the ratio \mbox{$\bar{J}/J=\lambda$}.

We have used ED on finite triangle chains with periodic boundary counditions consisting of up to $N_{\rm tr}=6$ triangles to study the ground-state phase diagram. We calculated the ground-state energy per site $\epsilon_0$ and the spin gap $\Delta$ as the difference between the ground states in the $S^z_{\rm tot}=1$ and $S^z_{\rm tot}=0$ sector. The corresponding data are shown as a function of $\lambda$ in Fig.~\ref{fig:triangle_chain_pbc} setting $J=1$. We find clear evidence for a breakdown of the trimerized phase around $\lambda_{\rm c}\approx 0.8$: i) the spin-gap scales to zero for this ratio and ii) the second derivative $\partial^2_\lambda \epsilon_0$ develops a resonance in the same $\lambda$-regime fully consistent with a divergence in the thermodynamic limit. Interestingly, we do not observe any further signatures for phase transitions in the range $\lambda>1$ which implies that the ground-state phase diagram of the triangle chain consists of two distinct quantum phases separated by a quantum phase transition at $\lambda_{\rm c}<1$. 

%
%Figure: GSE Triangular chain
%%%%%%%%%%%%%%%%%%%%%%%%%%%%%%%%%%%%%%%%%%%%%%%%%%%%%%%%%%%%%%%%%%%%%%%%%%%%%%%%%%%%%%%%%%%%%
\begin{figure}
\begin{center}
\epsfig{file=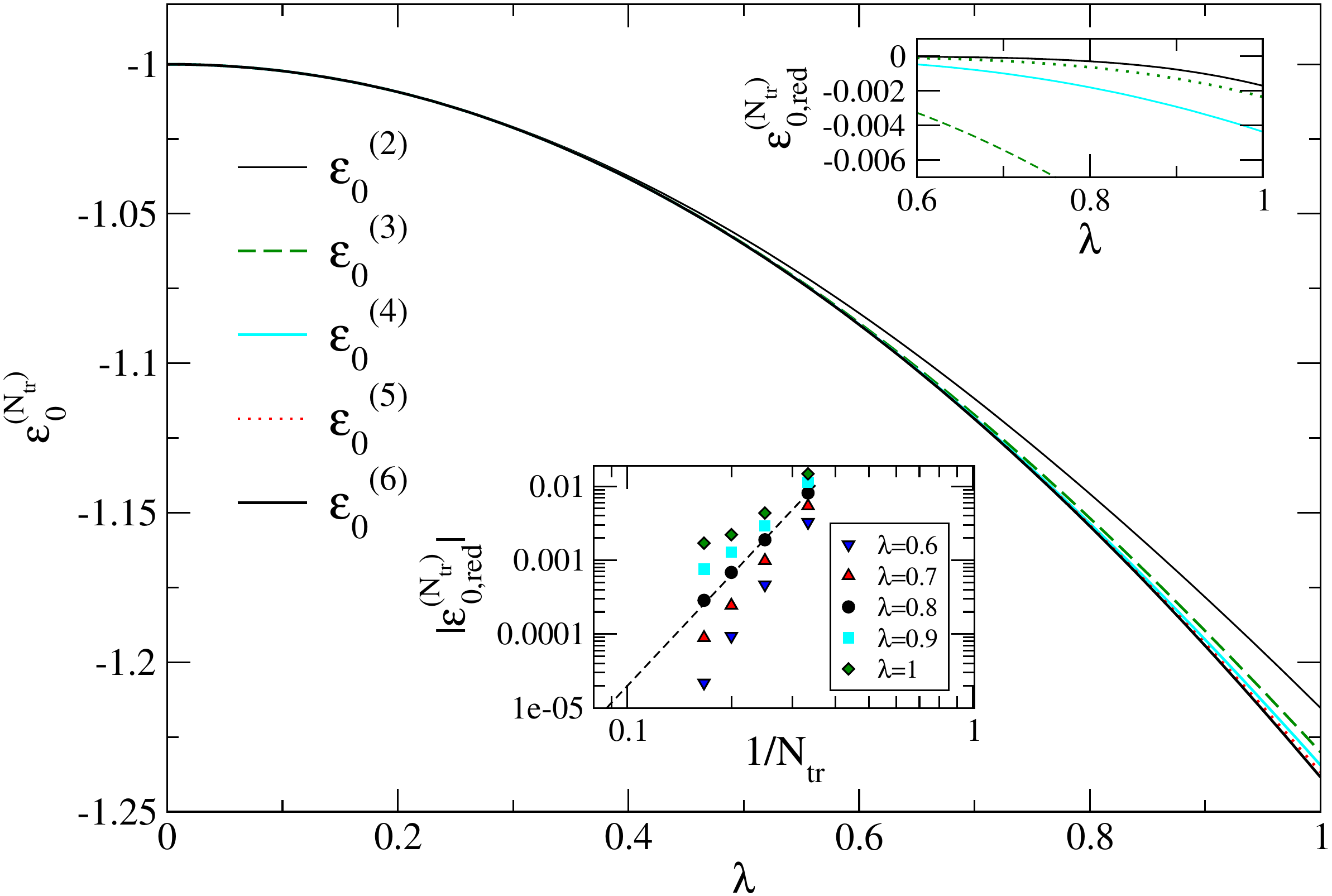, width=\columnwidth}
\end{center}
\caption{(Color online) Ground-state energy per site $\epsilon_{\rm 0}^{(N_{\rm tr})}$ of the one-dimensional triangle chain as a function of $\lambda$ calculated by NLCE using a full graph expansion for different $N_{\rm tr}$ and setting $J=1$. {\it Upper inset}: Reduced contributions $\epsilon_{\rm 0,red}^{(N_{\rm tr})}$ as a function of $\lambda$ for different $N_{\rm tr}$. {\it Lower inset}: Reduced contributions $|\epsilon_{\rm 0,red}^{(N_{\rm tr})}|$ as a function of $1/N_{\rm tr}$ for fixed values of $\lambda$ in a double logarithmic plot. Dashed black line represents an algebraic fit $a N_{\rm tr}^{-b}$ for the displayed data with $\lambda=0.8$.}
\label{fig:e0_triangle_chain}
\end{figure}
%%%%%%%%%%%%%%%%%%%%%%%%%%%%%%%%%%%%%%%%%%%%%%%%%%%%%%%%%%%%%%%%%%%%%%%%%%%%%%%%%%%%%%%%%%%%%%%

Next we turn to our NLCE results for the triangle chain. The NLCE recipe for one-dimensional systems becomes particularly simple. Graphs are linked chain segments containing $N_{\rm tr}$ up-triangles, so for each $N_{\rm tr}$ there is just one graph, and one has for all embedding factors $\nu^{\mathcal{G}_j}=1$. Furthermore, the NLCE ground-state energy per site in the thermodynamic limit can be expressed analytically as 
%%%%%%%%%%%%%%%%%%%%%%%%%%%%%%%%%%%%%%%%%%%%%%%%%%%%%%%%%%%%%%%%%%%%%%%%%%%%%%%%%%%%%%%%%%%%%
\begin{equation}
 \epsilon_0^{(N_{\rm tr})} = E_0 ^{\mathcal{G}_{N_{\rm tr}}} -  E_0 ^{\mathcal{G}_{N_{\rm tr}-1}}\quad ,
\label{eq:e0_chain}
\end{equation}
%%%%%%%%%%%%%%%%%%%%%%%%%%%%%%%%%%%%%%%%%%%%%%%%%%%%%%%%%%%%%%%%%%%%%%%%%%%%%%%%%%%%%%%%%%%%%
since all other subcluster contributions cancel out exactly. The corresponding results for $\epsilon_0^{(N_{\rm tr})}$ up to \mbox{$N_{\rm tr}=6$} are displayed in Fig.~\ref{fig:e0_triangle_chain}. One observes a smooth behaviour and a very good convergence for increasing values of $N_{\rm tr}$. This is even better seen in the reduced contributions \mbox{$\epsilon_{0,{\rm red}}^{(N_{\rm tr})}\equiv \epsilon_{0}^{(N_{\rm tr})}-\epsilon_{0}^{(N_{\rm tr}-1)}$} shown in the upper inset of Fig.~\ref{fig:e0_triangle_chain}. They get smaller with increasing $N_{\rm tr}$ and larger with increasing $\lambda$. No signatures from the quantum critical point around $\lambda_{\rm c}$ can be detected in these quantities (the same is true for $\partial^2_\lambda \epsilon_0$ which is not shown). This is very likely due to the fact that quantum fluctuations from the trimerized phase are optimally taken into account with the chosen clusters while corresponding fluctuations of the phase present for $\lambda>\lambda_{\rm c}$ are only captured deficiently.      

%
%Figure: GSE Kagome S1 / Comparison iPEPS - Full
%%%%%%%%%%%%%%%%%%%%%%%%%%%%%%%%%%%%%%%%%%%%%%%%%%%%%%%%%%%%%%%%%%%%%%%%%%%%%%%%%%%%%%%%%%%%%
\begin{figure}
\begin{center}
\epsfig{file=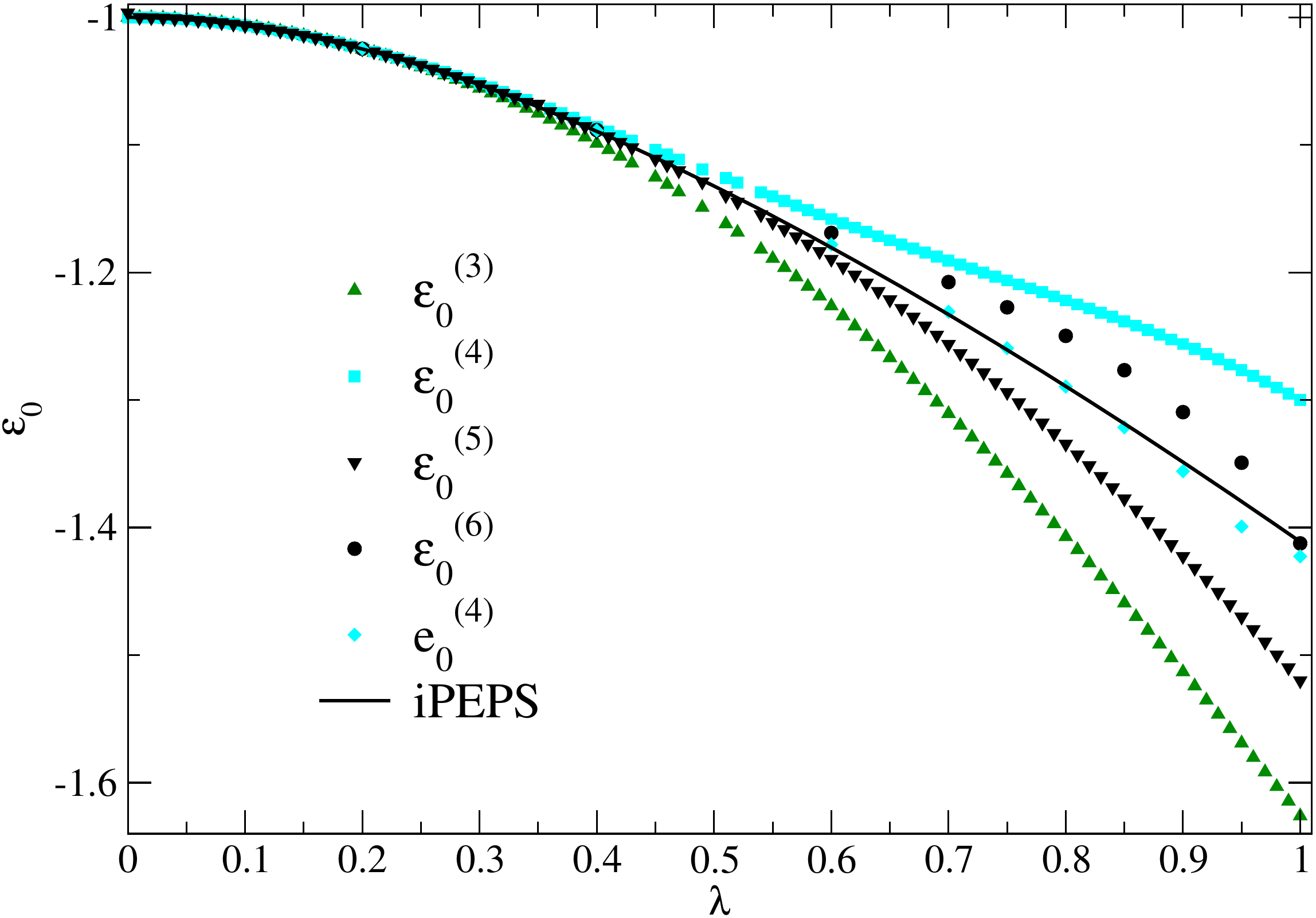, width=\columnwidth}
\end{center}
\caption{(Color online) Comparison of the ground-state energy per site $\epsilon_{\rm 0}$ as a function of $\lambda$ for $J=1$ calculated by NLCE using a full graph expansion for different $N_{\rm tr}$ (shown as symbols) and by iPEPS (shown as solid black line) taken from Ref.~\onlinecite{Liu15}. The gray/cyan diamonds correspond to the Wynn extrapolation $e_0^{(4)}$ where all contributions from $N_{\rm tr}=2$ to $N_{\rm tr}=6$ are taken into account.}
\label{fig:e0_full}
\end{figure}
%%%%%%%%%%%%%%%%%%%%%%%%%%%%%%%%%%%%%%%%%%%%%%%%%%%%%%%%%%%%%%%%%%%%%%%%%%%%%%%%%%%%%%%%%%%%%%%

Interestingly, there is an alternative way to grasp the quantum critical behaviour in the presented NLCE data for the triangle chain. In gapped phases one expects an exponential convergence of the NLCE data with increasing $N_{\rm tr}$ due to the finite correlation length which should be the case for all $\lambda$ except $\lambda_{\rm c}$. In contrast, at  $\lambda_{\rm c}$ (or close to  $\lambda_{\rm c}$ for any finite $N_{\rm tr}$) there should be an algebraic convergence due to the diverging correlation length when enlarging $N_{\rm tr}$ (see also Ref.~\onlinecite{yang11}). In the lower inset of Fig.~\ref{fig:e0_triangle_chain} the corresponding data of the reduced contributions $|\epsilon_{\rm 0,red}^{(N_{\rm tr})}|$ are displayed as a function of $1/N_{\rm tr}$ for several fixed values of $\lambda$ in a double logarithmic plot. Clearly, only the data set for \mbox{$\lambda=0.8\approx\lambda_{\rm c}$} is well described by an algebraic convergence and there are visible deviations for values of $\lambda$ being smaller or larger than $\lambda_{\rm c}$.

To summarize this part, we have found that the one-dimensional spin-one triangle chain possesses a quantum critical point around  $\lambda_{\rm c}\approx 0.8$, i.e.~the trimerized phase is not stable up to the isotropic value $\lambda=1$. As a consequence of this quantum critical behaviour, the reduced NLCE contributions of the ground-state energy per site only decay algebraically with cluster size close to $\lambda_{\rm c}$.

\subsection{Kagome lattice}
\label{SSect:results_kagome}

We start by discussing our NLCE results for the ground-state energy per site $\epsilon_0^{N_{\rm tr}}$ obtained from a full graph expansion in terms of triangles for the spin-one Heisenberg model on the kagome lattice. The corresponding data is shown together with infinite Projected Entangled Pair States (iPEPS) results taken from Ref.~\onlinecite{Liu15} in Fig.~\ref{fig:e0_full}. Compared to the case of the triangle chain discussed in the last section, the convergence behaviour is clearly not as good with increasing $N_{\rm tr}$. Curves with even and odd $N_{\rm tr}$ belong to two different families of data sets, since all curves with even (odd) $N_{\rm tr}$ are above (below) the iPEPS data. The NLCE curves converge very slowly for larger values of $\lambda$ and there are visible anomalies as a function of $\lambda$ already before the isotropic point $\lambda=1$. A certain improvement is achieved by applying the Wynn algorithm. In order to include the contribution of the highest order $N_{\rm tr}=6$, one has to exclude the lowest-order contribution as explained in Subsect.~\ref{SSect:extrapolation}. Therefore, the last step of the Wynn extrapolation yields $e_0^{(4)}$ which is also shown in Fig.~\ref{fig:e0_full}.  
%
%Figure: Kagome chain graphs - embedding factor - total contributions
%%%%%%%%%%%%%%%%%%%%%%%%%%%%%%%%%%%%%%%%%%%%%%%%%%%%%%%%%%%%%%%%%%%%%%%%%%%%%%%%%%%%%%%%%%%%%
\begin{figure}
\begin{center}
\epsfig{file=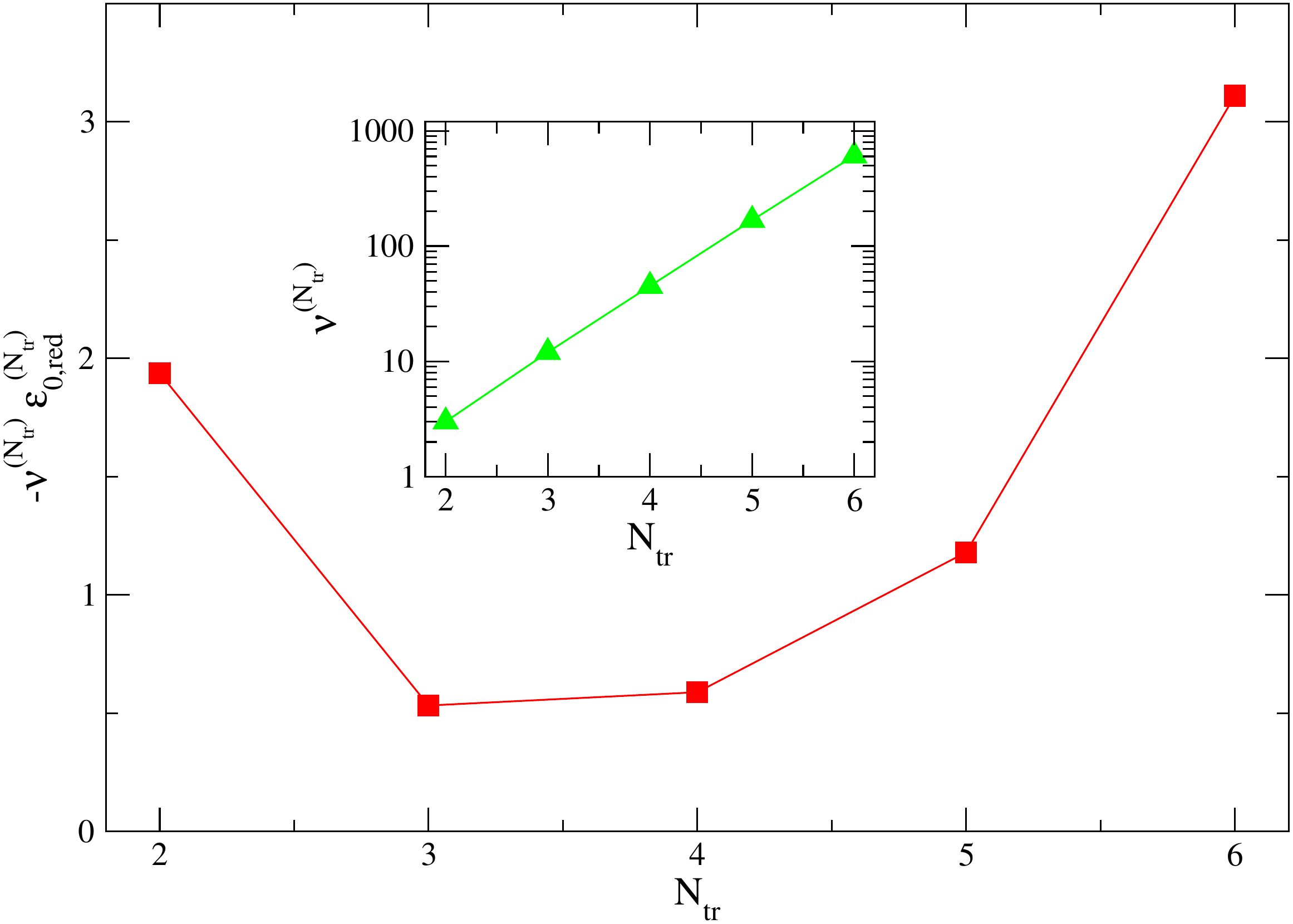, width=\columnwidth}
\end{center}
\caption{(Color online) Total contribution $-\nu^{N_{\rm tr}} \epsilon_{\rm 0,red}^{(N_{\rm tr})}$ of unfrustrated chain graphs as a function of $N_{\rm tr}$ for $\lambda=1$ where $\nu^{N_{\rm tr}}$ is the embedding factor and $\epsilon_{\rm 0,red}^{(N_{\rm tr})}$ the reduced contribution to the energy per site. {\it Lower inset}: Embedding factor $\nu^{N_{\rm tr}}$ of unfrustrated chain graphs as a function of $N_{\rm tr}$ for $\lambda=1$.}
\label{fig:kagome_chain_graphs}
\end{figure}
%%%%%%%%%%%%%%%%%%%%%%%%%%%%%%%%%%%%%%%%%%%%%%%%%%%%%%%%%%%%%%%%%%%%%%%%%%%%%%%%%%%%%%%%%%%%%%%   
%
%

This poor convergence is a consequence of the quantum critical behaviour of the unfrustrated triangle chain deduced in the last subsection. Indeed, all the triangle chain clusters of the one-dimensional triangle chain are also present in the full graph expansion in terms of triangles on the kagome lattice. These clusters are essentially unfrustrated and we therefore find that reduced contributions are negative. Physically, this means that these clusters lower the energy per site of the trimerized ground state, while the contributions of other clusters typically increase the energy due to the geometric frustration. We stressed in Subsect.~\ref{SSect:results_1d} that reduced contributions $\epsilon_{\rm 0,red}^{(N_{\rm tr})}$ of chain graphs decay only algebraically with $N_{\rm tr}$ close to $\lambda_{\rm c}\approx 0.8$. For the two-dimensional kagome lattice, this becomes problematic due to the fact that the embedding factor of these graphs grows {\it exponentially} with $N_{\rm tr}$. Consequently, the product $\nu^{N_{\rm tr}} \epsilon_{\rm 0,red}^{(N_{\rm tr})}$, which enters the NLCE expression for the ground-state energy per site Eq.~\ref{eq:e0}, {\it diverges} when enlarging the cluster size of the chain graphs (see also Fig.~\ref{fig:kagome_chain_graphs} for $\lambda=1$). This divergent partial series of the unfrustrated chain graphs has to be compensated by the overall contribution of the majority of remaining frustrated graphs to get a finite total result for the ground-state energy per site. Therefore, the poor convergence seen in Fig.~\ref{fig:e0_full} for the full graph expansion in terms of triangles should not come as a surprise when truncating the NLCE series by taking a finite $N_{\rm tr}$.

%
%Figure: GSE Kagome S1 / Comparison iPEPS - Full
%%%%%%%%%%%%%%%%%%%%%%%%%%%%%%%%%%%%%%%%%%%%%%%%%%%%%%%%%%%%%%%%%%%%%%%%%%%%%%%%%%%%%%%%%%%%%
\begin{figure}
\begin{center}
\epsfig{file=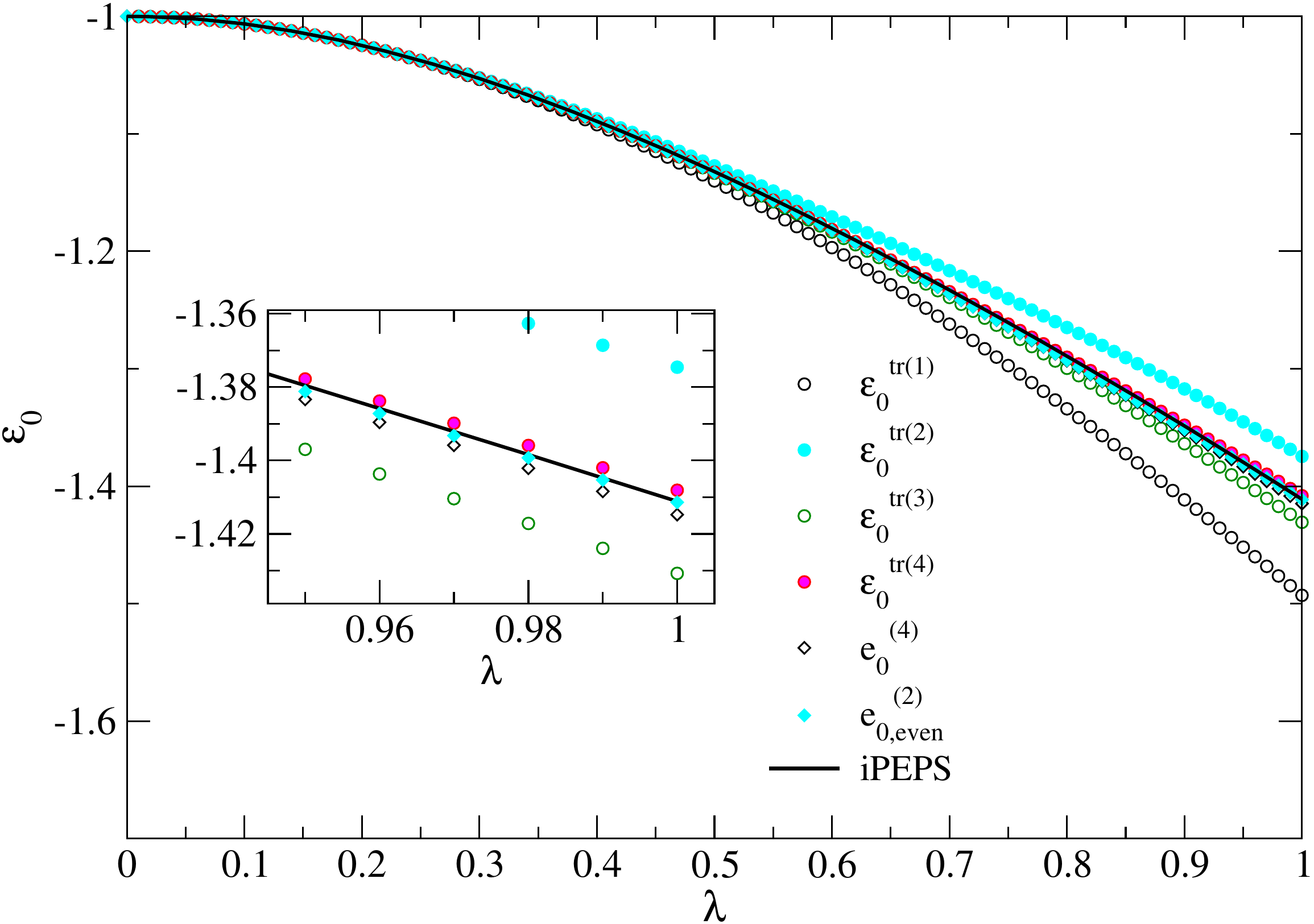, width=\columnwidth}
\end{center}
\caption{(Color online) Comparison of the ground-state energy per site $\epsilon_{\rm 0}$ for $J=1$ as a function of $\lambda$ calculated by NLCE using reorganized graphs of different size (shown as symbols) and by iPEPS (shown as solid black line) taken from Ref.~\onlinecite{Liu15}. Diamonds correspond to two different Wynn extrapolations. The open black (filled gray/cyan) diamonds belong to $e_0^{(4)}$ ($e_{0,\rm even}^{(2)}$) where all (even order) contributions are taken into account. {\it Inset}: Zoom of the same data close to $\lambda=1$.}
\label{fig:e0_triangle}
\end{figure}
%%%%%%%%%%%%%%%%%%%%%%%%%%%%%%%%%%%%%%%%%%%%%%%%%%%%%%%%%%%%%%%%%%%%%%%%%%%%%%%%%%%%%%%%%%%%%%%

One obvious idea is to reorganize the graph expansion in such a way that there are no divergencies in partial series resulting from quantum critical behaviour of one-dimensional unfrustrated subsystems. And indeed, using the two inequivalent effective triangles as frustrated and more symmetric building blocks, as introduced in Subsect.~\ref{SSect:graphs_opt}, the convergence of the obtained NLCE ground-state energy per site is dramatically improved as shown in Fig.~\ref{fig:e0_triangle}. Again, the results $\epsilon_0^{{\rm tr}(n)}$ alternate for $n$ being odd or even, but all curves smoothly approach the iPEPS result and absolute differences between NLCE and iPEPS are considerably smaller in the whole interval \mbox{$\lambda\in\left[0,1\right]$}.

Further improvement is achieved by applying the Wynn algorithm. In order to include the contribution of the highest order $N_{\rm tr}=4$, we include the contribution of a single supersite as the lowest-order contribution (see also Subsect.~\ref{SSect:extrapolation}). The last step of the Wynn extrapolation then yields $e_0^{(4)}$. For this extrapolation we find that the difference to the iPEPS data is similar compared to the one with the bare NLCE result $\epsilon_0^{\rm tr(4)}$ at $\lambda=1$ (but with a different sign). Interestingly, if we perform the Wynn algorithm restricting to even-order contributions, which corresponds to an effective plaquette expansion, the obtained extrapolant $e_{0,{\rm even}}^{(2)}$ shows almost perfect agreement with the existing numerical data even at $\lambda=1$: The value of the extrapolated iPEPS ground-state energy per site is $-1.4116(4) J$ from Ref.\onlinecite{Liu15}, the one calculated by DMRG from Ref.~\onlinecite{Changlani15} is $-1.410(2) J$ while we find $e_{0,{\rm even}}^{(2)}=-1.4114 J$.

\section{Conclusions}
\label{Sect:conclusions}
In this work we have studied the trimerized spin-one Heisenberg model on the kagome lattice applying non-perturbative linked-cluster expansions. To this end a parameter $\lambda$ is introduced which interpolates between the limit of isolated triangles and the isotropic Heisenberg model. We used a NLCE in terms of triangles for the ground-state energy per site which, by construction, becomes perturbatively controlled for small $\lambda$. 

This full graph expansion shows a non-monotonic convergence behaviour when approaching the isotropic Heisenberg model. Our findings can be traced back to quantum critical behaviour from unfrustrated quasi-1d chain graphs. For these graphs the embedding factor diverges faster (exponentially) with increasing graph size than the suppression of reduced contributions (algebraic behaviour) leading to an overall divergence of this partial subclass of graphs. Such divergencies have to be compensated by the remaining majority of frustrated graphs in order to yield a physically meaningful result which is a subtle issue when truncating the sequence in a NLCE calculation. The appearance of divergencies in partial series of unfrustrated graphs is likely a generic feature of NLCE calculations on frustrated systems using a full graph expansion.  

In our opinion there are two possibilities to overcome this finding. First, one can reorganize the graph expanion so that it is better suited for the problem under study. More concretely, one must avoid divergent partial series of graphs. For the kagome spin-one Heisenberg model we have shown that a reorganization of graphs in terms of effective triangles (hexagons and triangles in the original kagome lattice) leads to an enormous improvement in the convergence giving quantitative agreement with other recent approaches. It would also be interesting in future to investigate other reorganized graph expansions for this problem. 

Second, one might wonder whether there exist an optimized full graph expansion which goes beyond the paradigm of using ED on graphs as recently deduced for NLCEs when targetting excited states \cite{coester15}. It was shown in Ref.~\onlinecite{coester15} that graph-based continuous unitary transformations (gCUTs) are capable to acount for the reduced symmetry of graphs which leads to artificial processes on graphs which are not present in the thermodynamic limit. It would be fascinating if a similar progress can be found for ground-state properties of frustrated systems.

Finally, it would be interesting to determine the properties of excited states in the spin-one kagome Heisenberg model which we leave for future studies. 

\begin{acknowledgments}
We thank K. Coester, W. Li, A. Weichselbaum, and J. von Delft for fruitful discussions as well as  W. Li for providing numerical data. This research has been suppoarted by the Virtual Institute VI-521 of the Helmholtz association.
\end{acknowledgments}

\end{document}